\documentclass[10pt,conference]{IEEEtran}
\usepackage{eurosym}
\usepackage{amssymb}
\usepackage{amsmath}
\usepackage{amsthm}
\usepackage{paralist}
\usepackage{graphicx}
\usepackage[font=footnotesize, justification=centering, labelsep=period]{caption}
\usepackage [noadjust]{cite}
\newcommand{\comment}[1]{}

\begin{document}
\title{\textbf{\Large Information Integration using the Typed Graph Model}}
\author{
\IEEEauthorblockN{Fritz Laux}
\IEEEauthorblockA{Fakult\"{a}t Informatik\\
Reutlingen University\\
D-72762 Reutlingen, Germany\\
email: fritz.laux@fh-reutlingen.de}
\and
\IEEEauthorblockN{Malcolm Crowe}
\IEEEauthorblockA{School of Computing\\ %, Engineering and Physical Sciences\\
University of the West of Scotland\\
Paisley PA1 2BE, UK\\
email: malcolm.crowe@uws.ac.uk}
}

\maketitle
\begin{abstract}
Schema and data integration have been a challenge for more than 40 years. 
While data warehouse technologies are quite a success story, there is still a lack of information integration methods, especially if the data sources are based on different data models or do not have a schema. 
Enterprise Information Integration has to deal with heterogeneous data sources and requires up-to-date high-quality information to provide a reliable basis for analysis and decision making. 
The paper proposes virtual integration using the Typed Graph Model to support schema mediation.
The integration process first converts the structure of each source into a typed graph schema, which is then matched to the mediated schema. 
Mapping rules define transformations between the schemata to reconcile semantics.
The mapping can be visually validated by experts.
It provides indicators and rules to achieve a consistent schema mapping, which leads to high data integrity and quality. 

\end{abstract}

\IEEEpeerreviewmaketitle

\section{Introduction}
\label{sec:intro}
Information integration, integrating data from different sources, enables us to gain new knowledge and insights that help to make predictive analysis, coordinate complex processes and control systems. 
Depending on the application the mediated data can be materialized as in a Data Warehouse (DW) or each query can access the data sources to get an up-to-date result, which is called \emph{virtual integration}. 
Many situations need up-to-date or near real-time information, which can only be achieved by virtual integration.
This is the case not only in emergency situations like fighting epidemic, earthquake, and other disaster aid but also in industry production. 

In Enterprise Information Integration (EII), the integration of heterogeneous data sources is usually achieved by a manually supervised process to ensure a high-quality mediated global schema.
In this paper, we concentrate on \emph{supervised} schema integration, i.e., the semantic data integration problem. 
Much research has been conducted to automatically match and map data sources \cite{Rahm}\cite{Melnik}, but the quality of the results are usually not sufficient for EII systems \cite{Haas, Bernstein2008, Golshan, Strong}. 
Thus, additional manual changes using context and other semantic information are necessary to build a high-quality global/mediated schema, i.e., an \emph{engineered schema} \cite{Bernstein2007}.

The manual improvement and quality checking of the mediated schema should be supported by software that visualizes the semantics of the data integration and the impact and interplay of any changes. 

\begin{figure}[]
\centering
\includegraphics[width=0.48\textwidth]{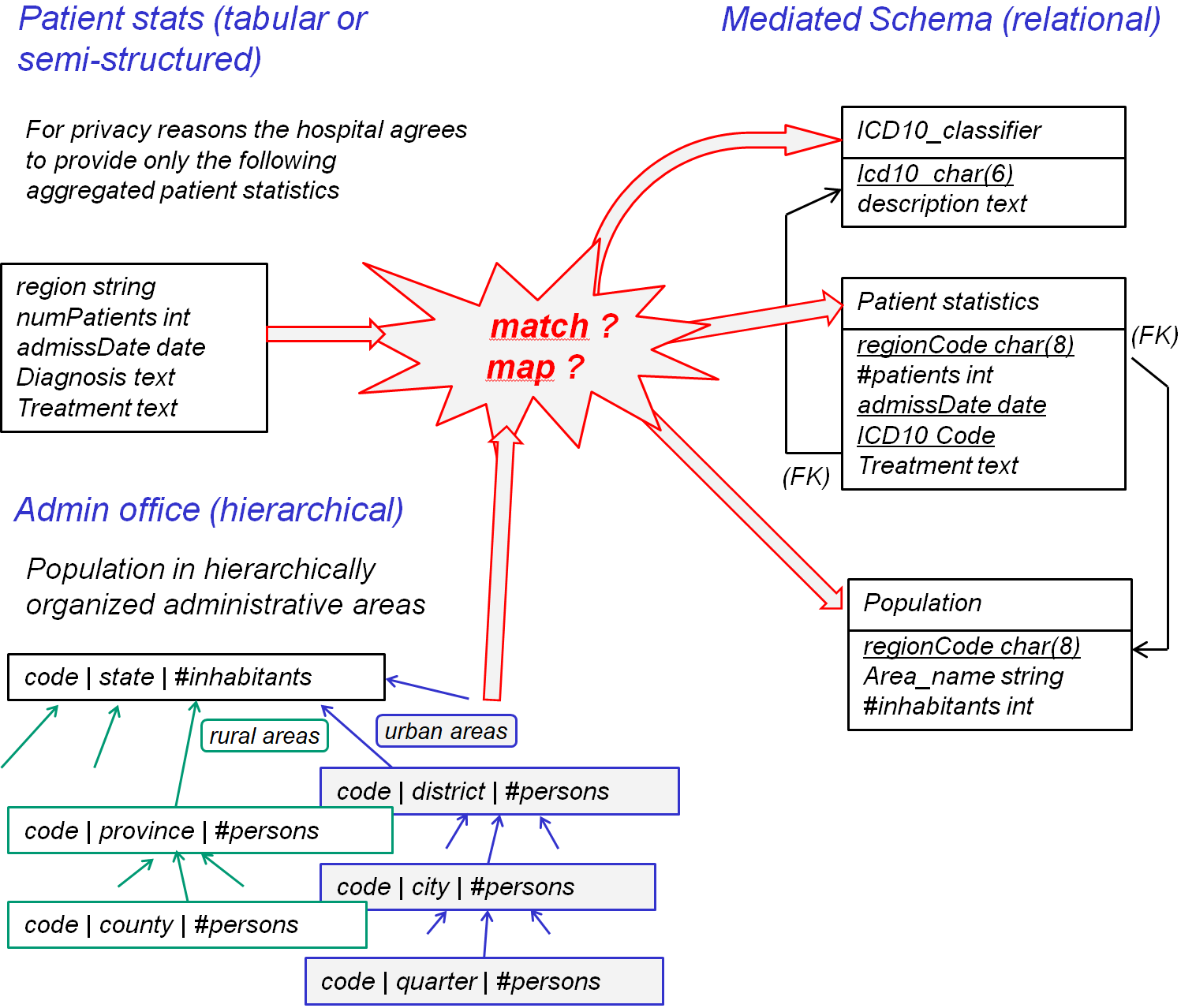}
\caption{Data integration scenario (Running example)}
\label{fig:RunningExample}
\end{figure}

Let us illustrate our approach with an example scenario presented in Figure \ref{fig:RunningExample}. It is borrowed from Crowe et al. \cite{Crowe}. 
The World Health Organization reports on emergencies/epidemics, using data provided in many different formats by national or regional autonomous actors like national health authorities or hospitals. 
For simplicity, we only show one hospital providing aggregated \textit{tabular data} on patients grouped by regions, admission date and diagnosis.
For each group element the number of patients and their treatment are recorded.
The statistics admin offices deliver demographic data in a \textit{hierarchical structure} reflecting the administrative areas.
We want to create an integrated/mediated schema that combines these sources using the International Classifier for Diseases (ICD).
Given a \textit{relational} Integration Schema, the question is ``how do we find matches for the data items and transform them".

We can identify five possible problems in this example.

From Figure \ref{fig:RunningExample}, we need to identify the matching between \textit{region string} of the hospital patient stats and the \textit{regionCode} in the mediated schema despite the different coding (Problem 1). 
There might be different spelling or naming of \textit{region} strings in different hospitals. 
These conflicts need to be resolved in the mediated schema (Problem 2). 
Preserving the semantics (Problem 3) requires the \textit{ICD10\_classifier} for the mediated schema to be obtained from the \textit{diagnosis} text. 
The population data from administrative areas need to be mapped into relations preserving the hierarchy (Problem 4). 
And finally (Problem 5), we need consistency when multiple mapping paths exist like from \textit{region} (hospital stats) to \textit{regionCode} (Population), one via \textit{code} (admin office) and the other either direct or via
\textit{regionCode} (Patient statistics). 

In Figure \ref{fig:RunningExample}, we used different graphical representations for the schemata to reflect the various data models and structures. 
But a consistent graphical representation would be better to solve the matching/mapping task. 
The Typed Graph Model (TGM) \cite{Laux} represents a flexible model with enough expressive power  to cover most data structures comprehensively and support several abstraction levels. 
This is why we have chosen the TGM for the schema mediation and data mapping.

Data integration has been intensively investigated from a theoretical point of view \cite{Doan, Lenzerini, Fagin2012} leading to mainly two concepts, \emph{Global As View} (GAV) and \emph{Local As View} (LAV). 
As the terms indicate the global (resp. local) data are expressed as view of local (resp. global) data. The ACM digital library alone retrieves 163 matches for the key words ``GAV or LAV". 
These papers are of great value to understand and specify the correspondences between two or more databases. 
But using description logic as formal specification gives little help to identify synonyms and homonyms or discover their semantics.
Where sources of low quality overlap, it is important to remove redundancy while seeking to benefit from any extra information. 
The mapping of heterogeneous data sources to a specific target does not allow a fully automatic procedure if high data quality is required 
\cite{Strong}\cite{Bernstein2008}\cite{Halevy2005}\cite{Maluf}.
% To deal with the listed questions we use the TGM to identify and map data elements.

\subsection{Contribution}
Our idea is to use the TGM to support, formalize, and visualize data integration. 
The TGM helps to create a mediated target schema, in contrast to the less formalized Extract-Transform-Load process used in Data Warehousing. 
We propose a process, which divides the integration task into two phases:
first, model all sources and the target using TGM, then, second, match and map the source models into the target model.
The proposed process is well defined and combines for the first time \emph{supervised} semi-automatic matching with mapping and merging of data.
It provides rules for mapping and conflict resolution and defines criteria for quality control.
We illustrate all integration steps using our running example.

\subsection{Structure of the Paper}
Section \ref{sec:RelatedWork} briefly presents theoretical data integration work and practical experiences with emphasis on high-quality integration.
In Section \ref{sec:DIFramework} the integration framework is presented. It consists of four activity steps: 
\begin{enumerate}
\item Modeling the source data structures with TGM (Subsection \ref{ssec:ModelSource})
\item Defining the target schema using TGM (Subsection \ref{ssec:ModelTarget})
\item Match/Map source with target data, resolve conflicts, and define necessary transformation (Subsection \ref{ssec:Transform})
\item Check and improve quality (Section \ref{sec:Quality}).
\end{enumerate}
Quality criteria and measures are developed in Section \ref{sec:Quality} to help check and improve the integration quality. 
The paper ends with a summary of our findings and gives an outlook on ideas for future work.

\section{Related Work}
\label{sec:RelatedWork}
Since the middle of 1980s many papers on data integration have been published. 
In the following review, we restrict the focus to high-quality integration targeting EII.

\subsection{Data Integration Overview}

The papers of Sheth and Larson \cite{Sheth} give an introduction to federated database systems. 
Later textbooks of \"Oszu and Valduriez \cite{Ozsu} and also Leser and Naumann \cite{Leser} describe different integration methods and present general approaches for schema matching and mapping. 
Because our focus is on high-quality integration the experiences and considerations of Bernstein/Haas \cite{Bernstein2008} and Halevy et al. \cite{Halevy2005} with real live projects are of high value to us. 
Both papers emphasize the need to integrate heterogeneous data sources including email, order documents, warranties, and other un- or semi-structured data. 

Laura Haas states in her paper \cite{Haas} that ``despite the weighty body of
literature, the information integration challenge is far from solved, especially in the
enterprise context". 
As ``Big I" challenges she identifies \emph{entity resolution} and the lack of \emph{theoretical and practical guidance} to make schema integration choices, noting the lack of a ``broader framework" with quality control, which ``considers the entire end-to-end integration process". 
This statement is confirmed by many papers that address the integration of Web data \cite{Friedman, Roman, Popa}, XML data \cite{Popa}, or RDF data \cite{Langegger}\cite{Adamou2020} in a declarative way but with little help on how to proceed in practice.

\subsection{Data Integration using Graph Models} 

Some authors use a Graph Data Model (GDM) for data analysis and transformation. 
G\textsc{RADOOP} \cite{Junghanns2015} and Pregel \cite{Malewicz} are example prototypes for this approach. 
%The SemWIQ Middleware \cite{Langegger} for instance has wrappers for relational databases (D2R), CSV files, and SOAP endpoints. 
Most of the integration of graph databases is instance level based and uses graph transformations, see \cite{Kricke}\cite{Melnik}, but the question of how well the integration matches the original semantics and schemata of the sources is not addressed.  
% Adamou \cite{Adamou2017} used just-in-time compilation for queries on linked data and he also used a relaxed GAV to mediate linked data \cite{Adamou2020}.  

None of the papers really addresses how to match and map differently structured data elements by preserving the semantics of the sources.
Practical guidance is available from de Sousa and del Val Cura \cite{DeSousa} only for the mapping from the Extended Binary Entity Relationship (EB-ER) model to a Property Graph Model (PGM). 

The only publication we are aware of that tries to improve data integration quality is Gelman \cite{Gelman}. 
In his paper he develops a theory that helps to produce accurate data integration output from multiple, overlapping, and inaccurate sources. 
He assumes that errors are not random and that ``complementary" information helps to select the most accurate data. 
This approach could also help with the entity resolution problem.

Another problem with virtual data integration is that the global database may not be consistent with respect to integrity constraints. 
Bertossi and Bravo \cite{Bertossi} recommend querying only the consistent part of the global database. 
They define a consistent answer to a query when the result is the answer to every ``repair" of the global database, i.e., a maximal subset of the global database that satisfies the integrity constraints.  
The solution to the problem is quite general and conceptually clear, however an implementation is still missing. 
The authors demand that these issues must be ``addressed in order to use those solutions in real database applications".

\section{The Integration Framework}
\label{sec:DIFramework} 
 
The data integration framework uses the TGM as intermediate data model. This is why the TGM is shortly presented here. 
A detailed and formal introduction can be found in \cite{Laux}. 
The data integration process consists of transforming the sources and target into Typed Graph Schemata (TGS) and then make the matching and mapping of schema elements.
 
\subsection{The Typed Graph Model}
\label{ssec:TGM}

The TGM combines schema support, complex data structures and abstraction using sub-graphs. 
Because of its rich semantics, we use it to capture the source and target semantics as accurately as possible. The source and target schemata act as a means of quality control. 
 
The TGM informally constitutes a directed property hyper-graph that conforms to a schema. 
Formally, the TGM is defined as quadruple $TGM = (N, E, TGS, \phi)$ where:
\begin{itemize}
\item
$N$ is the set of named (labeled) nodes $n$ with data types from $N_S$ of schema TGS.
\item
$E$ is the set of named (labeled) edges $e$ with properties of types from $E_S$ of schema TGS.
\item
$TGS$ is a typed graph schema defined as tuple $TGS = (N_S, E_S, \rho, T, \tau, C)$, where $N_S, E_S$ are vertices and edges of the graph schema, $\rho$ defines the hyper-edges with its cardinalities $\tau$. 
$T$ is a set of data types used for vertices and edges, and finally $C$ is a set of integrity constraints, which the graph database must obey.
\item
$\phi$ is a homomorphism that maps each node $n$ and edge $e$ of $TGM$ to the corresponding type element of $TGS$,
\end{itemize}

As graphical representation for the TGS, we adopt the Unified Modeling Language (UML) to visualize the data model, which makes it useful for visualization tools to support the human controlled information matching and mapping.  
The visualization allows the modeling expert to validate the matching and mapping. 
The possible abstraction via sub-graphs facilitates the overview and management of complex and large models.

A data model can act as a kind of \textbf{supermodel} if it is general enough to capture all popular models. 
Hull \cite{Hull} and Atzeni et al. \cite{Atzeni1996}\cite{AtzeniVLDB08} describe such a supermodel  that subsumes all popular data models including the Relational Model (RM), ERM, XML, Object Oriented Model (OOM), Object Relational (OR), and XSD. 
It consists of the following meta-constructs: \emph{lexical, abstract, aggregation, generalization,} and \emph{function}. 
The TGM can represent these meta-constructs in an information preserving way \cite{Laux2021} by matching one-to-one lexical elements with properties, abstract constructs with nodes, aggregation and generalization constructs with the corresponding edge types, and functions with directed edges having multiplicity 1. 
This implies that the TGM is able to map the above-mentioned data models without loss of any information.
% The matching can be chained or composed for complex composites. 

% The UML class symbol is used for nodes and include its properties as attributes including their data type. Labels are written in the top compartment of the UML-class. Edges of the TGS are represented by UML associations. For the label and properties of an edge we use the UML-association class, which has the same rendering as an ordinary class but its existence depends on an association (edge), which is indicated by a dotted line from the association class to the edge. This not only allows to label an edge but to define user defined edge types.

\subsection{The Data Integration Process}
\label{ssec:DIProcess}
 
\begin{figure*}[]
\centering
\includegraphics[width=0.85\textwidth]{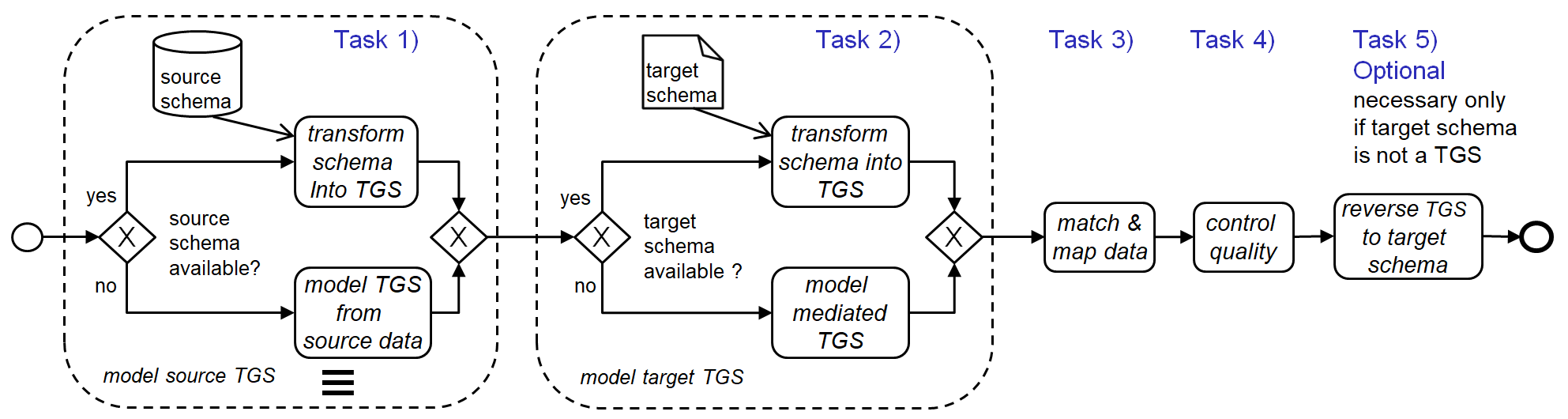}
\caption{Workflow of information integration using the TGM}
\label{fig:Workflow}
\end{figure*}

The benefit of information integration is maximized when source data are integrated with their full semantics. We believe a key success factor is to model the sources and target information as accurately as possible. The expressive power and flexibility of the TGM allows to describe the meta-data of the sources and target precisely and in the same model, which simplifies the matching and mapping of the sources to the target.  

The data integration process consists of five tasks:
\begin{enumerate}
\item model sources as TGS $S_i \quad (i = 1, 2, ..., n)$
\item model target schema $T$ as TGS $G$
\item match and map sources $S_i$ with TGS $G$
\item check and improve quality
\item convert TGS $G$ back to $T$ again
\end{enumerate}
The full process is depicted in Figure \ref{fig:Workflow} as a workflow modeled in BPMN. 
The $4^{th}$ task of the workflow is crucial for EII and other data integration projects, which demand highly accurate information quality. 
It might turn out that the resulting quality is not sufficient. 
As consequence the process might have to be iterated with different mappings in order to improve the quality.
The last task is only necessary if the target schema is not a graph schema.

The first and second tasks consist of two alternatives depending on the pre-existence of source schemata, resp. target schema. 
If a schema already exists it is only necessary to transform it into a TGS. 
If a source has no schema, it is then necessary to collect structure and type information from a data expert or from additional information. 
This information is necessary before an appropriate TGS can be created. 
At least a minimal schema is required for every data source. 
This extra effort has the advantage to ensure a better data quality.

\subsection{Model the Data Sources as TGS (Task 1)}
\label{ssec:ModelSource}

The relevant data must first be identified together with its meta-data if available. 
This includes coding and names for the data items. 
The measure units and other meta-data provided by the data owner are used to adjust all measures to the same scale.

If the source is a database or other rigid structured data, the modeling of a TGS is rather simple and there are publications \cite{DeVirgilio}\cite{Stoica} that propose automatic schema conversion. 
The paper of Laux \cite{Laux} gives some examples how to transform relational, object oriented, and XML-schemata into a TGS. 
If a schema already exists for a data source it may seem to be extra work to model it again as TGS. 
The benefit comes later when we look for matches because the schemata are all based on the same model. 
In addition, the quality of the matching can be checked formally with graph analysis. 
%In case of autonomous sources the owner may not want to disclose all data. He usually provides only a partial view on the data.

If the source is unstructured or semi-structured, e.g., documents or XML/HTML data, concepts and mechanisms from Information Retrieval (IR) and statistical analysis may help to identify some implicit structure or identify outliers and other susceptible data. 
If the data are self-describing (JSON, key-value pairs, or XML) linguistic matching can be applied with additional help from a thesaurus or ontology. 
Nevertheless, it is advisable to validate the matching with instance data or an information expert.

\begin{figure}[]
\centering
\includegraphics[width=0.48\textwidth]{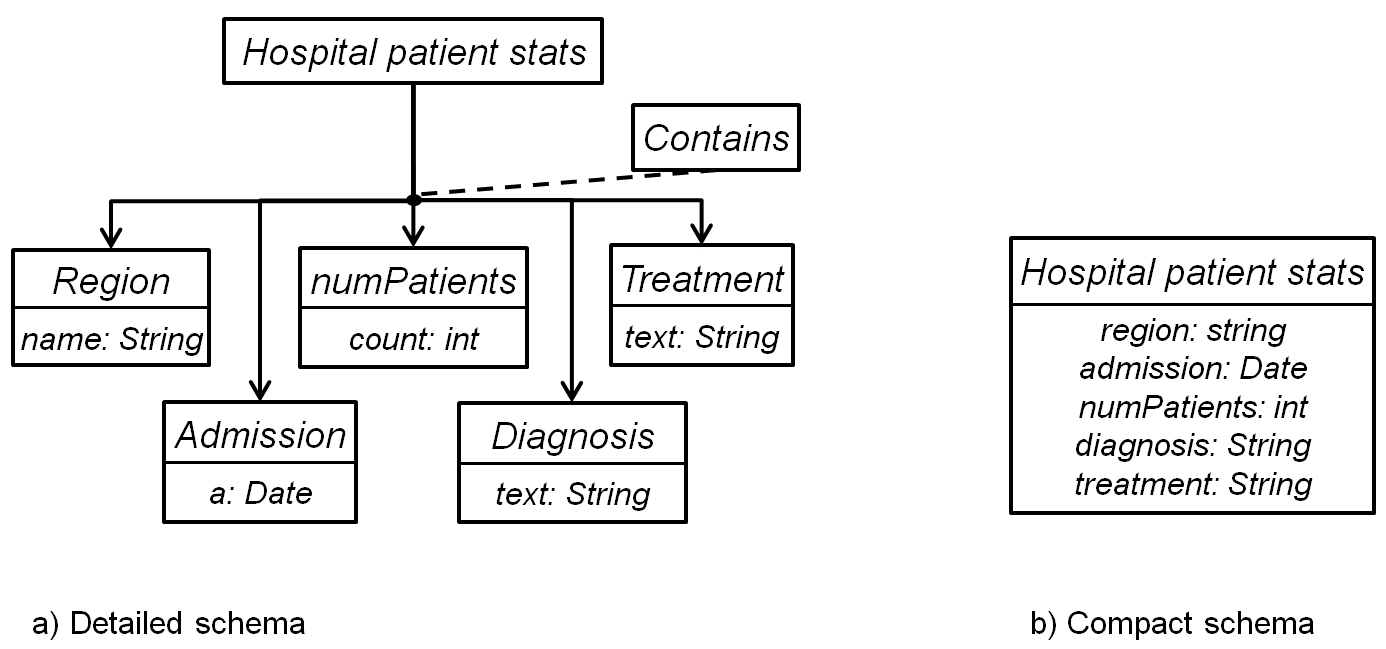}
\caption{TGS in UML notation for the patient statistics}
\label{fig:PatientStats}
\end{figure}

As example for semi-structured source data, we take the hospital patient statistics from our running example. 
The data for the statistics are entered in a form. 
We present two possible TGS in UML notation in Figure \ref{fig:PatientStats}. 
Part a) shows a detailed schema where each data element is modeled as a node with its corresponding property and (simple) data type. 
% The whole patient statistics could be modeled as one node with all properties contained (part b). 
In part b) the hospital patient stats are modeled as a complex data type. 
This little example demonstrates already the flexibility of the model in terms of detail and abstraction. 

\subsection{Model Mediated TGS (Task 2) and Target Schema (Task 5)}
\label{ssec:ModelTarget}

In general, many integration schemata are possible. In most cases the mediated schema tends to cover the union of the source schemata.
Petermann et al. \cite{PetermannVLDB} present an automatic graph instance integration with the help of a Unified Metadata Graph (UMG). 
This method can be applied to generate a mediated graph TGS if the UMG is replaced by the data source schema graphs. 
Another approach for unstructured data is proposed by Buneman et al. \cite{Buneman} by union of the source graphs. 

If a target schema $T$ is given, but not already as TGS, it needs to be transformed to a TGS. 
In most cases a 1--1 transformation is possible.
This can be done automatically using the same techniques as mentioned in \ref{ssec:ModelSource}.
This is the only case where Task 5) has to be executed. 
The TGS has to be reversed in this case to the target schema $T$ again by applying the inverse transformation.

\begin{figure}[]
\centering
\includegraphics[width=0.48\textwidth]{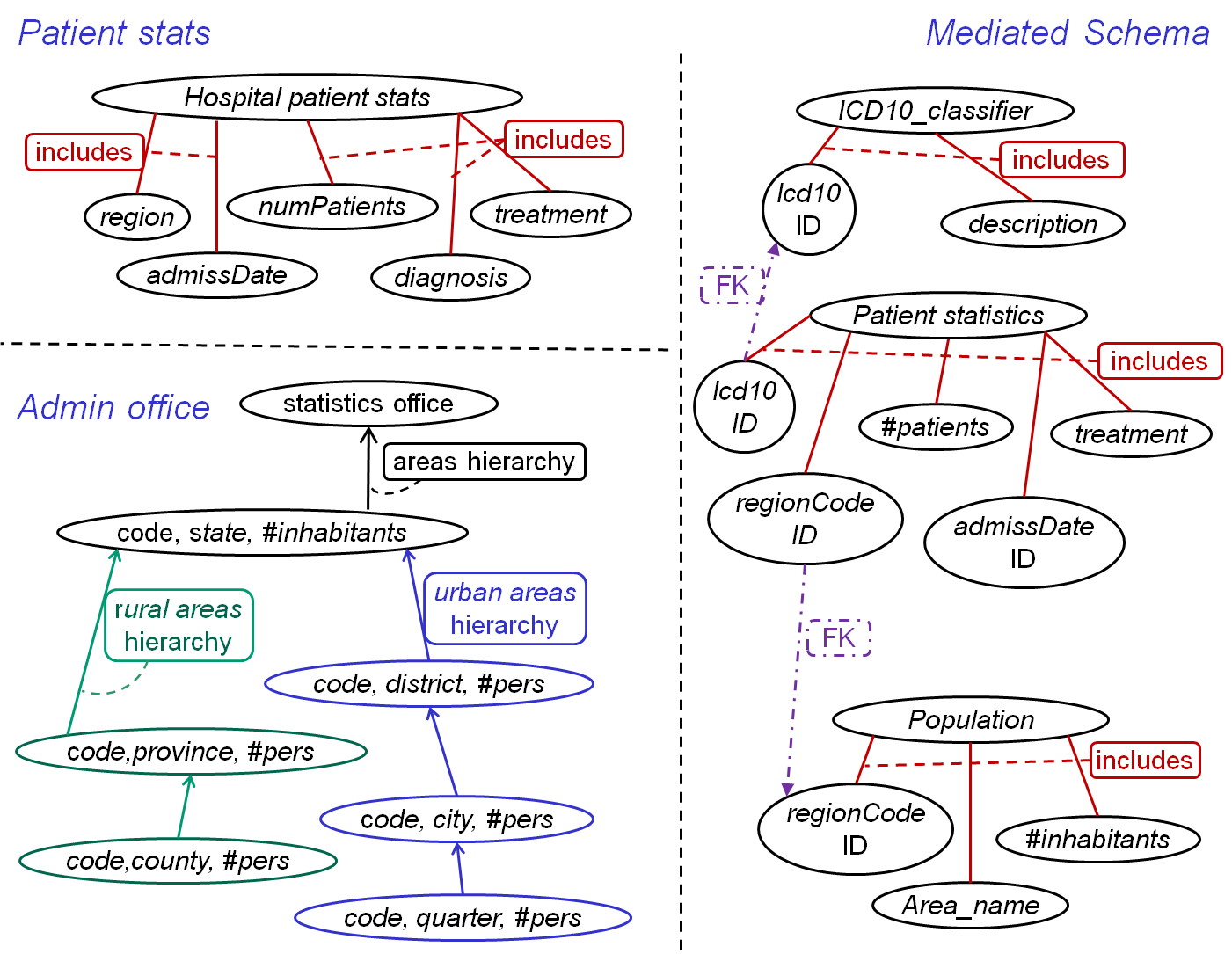}
\caption{Running example as TGS (data types not shown for simplicity)}
\label{fig:RunningExample2}
\end{figure}

Let us return to our running example and model the two data sources and the mediated target as TGS. 
The result is shown in Figure \ref{fig:RunningExample2}. 
The nodes of the schemata are visualized with property names inside the ovals. 
The data types of the properties are suppressed, but the edge types (labels) are color coded and connected by dashed lines with the corresponding edges.  
In the next task, we have to decide on the matching and mapping of vertices.

\subsection{Matching \& Mapping Source TGS to Target TGS (Task 3)}
\label{ssec:Transform}

%With the three TGS at hand we can start to match nodes and edges.
Task 3 addresses problem 1 from the Introduction \ref{sec:intro}.   
The matching step only identifies nodes and edges that are related, not necessarily one-to-one. 
In our example hospital statistics, apart from national language differences, the English names: cases, positive cases, reported cases, hospitalized, etc. could mean all the same or could mean different things.
The TGM % -– a labeled property graph model enhanced with data types - 
can help to identify, visualize, and match nodes and edges correctly using type and edge information (edge type, structure analysis, description). 
The matching can be supported by linguistic methods (name similarity, synonym and homonym list, thesaurus or ontology) and value analysis, e.g., using duplicates \cite{Bilke}. 
There are some publications that automate this process, see \cite{Rahm}\cite{Melnik}.
The use of TGM is flexible enough to support various data structures and visualizes the integration process, which helps to identify and resolve mapping problems manually.

In the next step, we define mappings between the source and the target nodes. 
The mapping is mainly a manual task and the integration schema designer has the responsibility to choose the best quality (freshness, reliability, precision) data, resolve conflicts if redundant data are available and to correct apparently incorrect names. 
These include merge operations with conflict resolution (Problem 2), e.g., entity deduplication, overlapping conflicting data, identification of global data, and (data type) transformation. 
It is important to preserve the semantics as far as possible (Problem 3). This requires the knowledge of meta-data (data type, structure analysis, description), which is supported by the TGM.
The mapping may include data grouping, e.g., grouping patients according to cost factors (ABC analysis).
The workflow sequence and the mapping function suggest following the more natural Global-As-View (GAV) approach when implementing the mappings. 
Even if the integration of new sources more complicated compared to the Local-As-View (LAV) approach, it reflects the reality of overlapping data with different coding or semantics, which has to be resolved in both approaches.

\begin{figure*}[]
\centering
\includegraphics[width=0.70\textwidth]{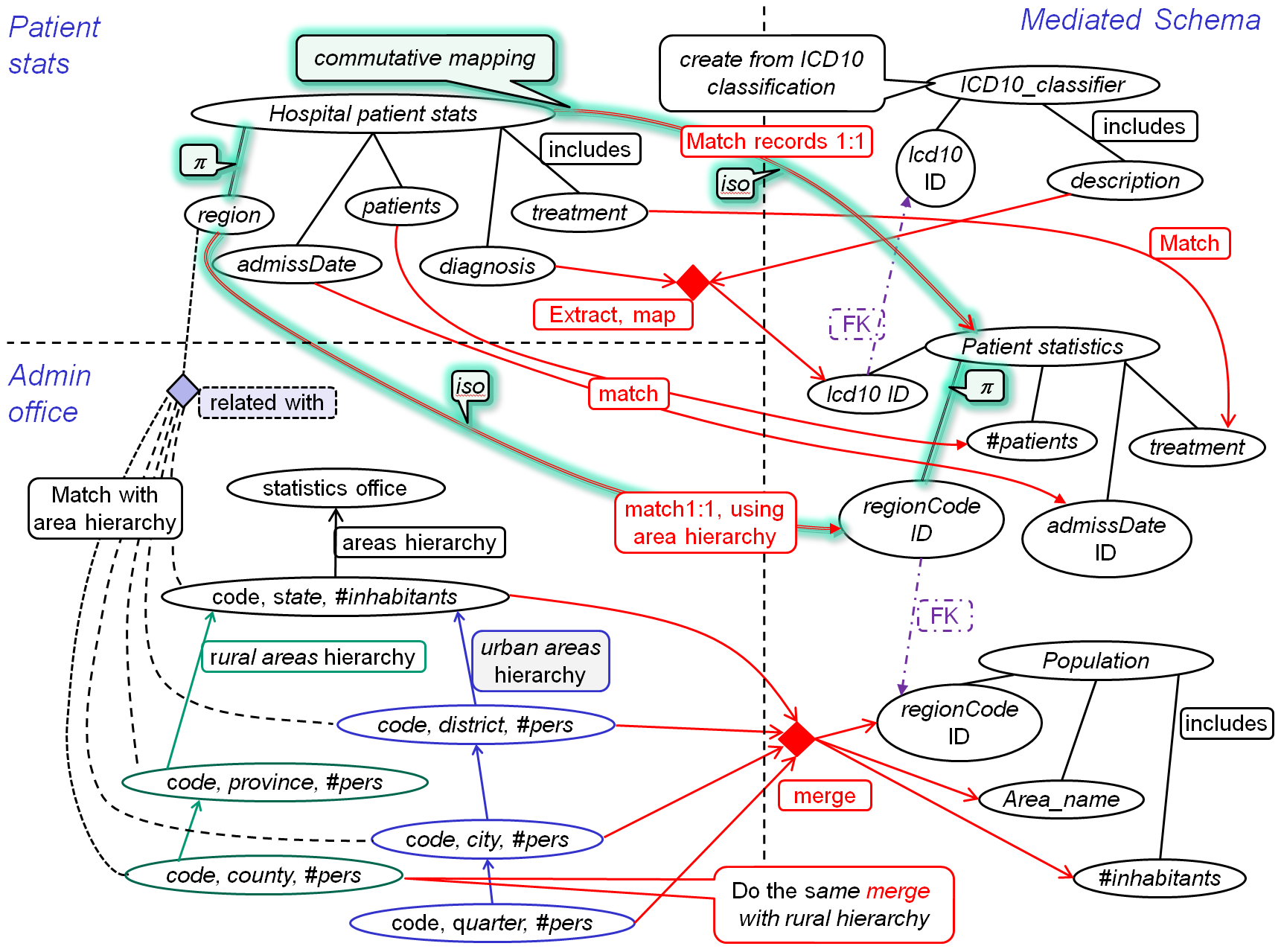}
\caption{Matching and mapping result of the complete running example. To not overload the figure, only some of the matches and mappings (red arrows) are shown. The commutative mapping between \textit{Hospital patient stats} and \textit{regionCode} is highlighted (green glow).}
\label{fig:Mapping}
\end{figure*}

To complete our running example, we present in Figure \ref{fig:Mapping} the result of the mapping of our running example. 
As in Figure \ref{fig:RunningExample2} on the left side are two TGS representing the data sources and on the right side is the \emph{Mediated Schema} which can be converted back to the relational target schema given in Figure \ref{fig:RunningExample}. 
This corresponds to the final step 5 of the workflow. 
For illustration purpose the Foreign Keys (FK) are indicated by dashed blue lines.
The two data source graphs are semantically connected (\emph{related with}) via the region information.
The matching and mappings (red arrows) between source schemata and target TGS exemplify only some of necessary matching and mappings.
Between \textit{Hospital patient stats} (source Patient stats) and \textit{regionCode} (mediated schema) exist two commutative mappings: (1) via \textit{Patient statistics} and (2) via \textit{region}. Both include an isomorphism (iso) and a projection ($\pi$).

\subsection{Example Patterns}
\label{ssec:Examples}

In order to illustrate the modeling power and flexibility of the TGM, we present a series of typical mappings that arise during schema integration and mediation. 
Such mapping patterns reoccur often and have a standardized solution.

\subsubsection{Merge Pattern}
The Merge pattern solves problem 2) of how to merge two or more data nodes of similar semantics.
The different data sources can provide additional and overlapping data.
Multiple sources may produce conflicting values or duplicates, differ in scale and coding, and have different resolution.
Duplicates must be removed. Rules need to be established for conflicting values, e.g., prioritize the most reliable value or calculate a mean value if all sources are of similar quality. 
In case of different coding use translation tables. 
For scale and unit mismatches define value transformations. 
The merge pattern is useful to reconcile data from different sources and to improve data quality if the data is redundant.

\begin{figure}[]
\centering
\includegraphics[width=0.48\textwidth]{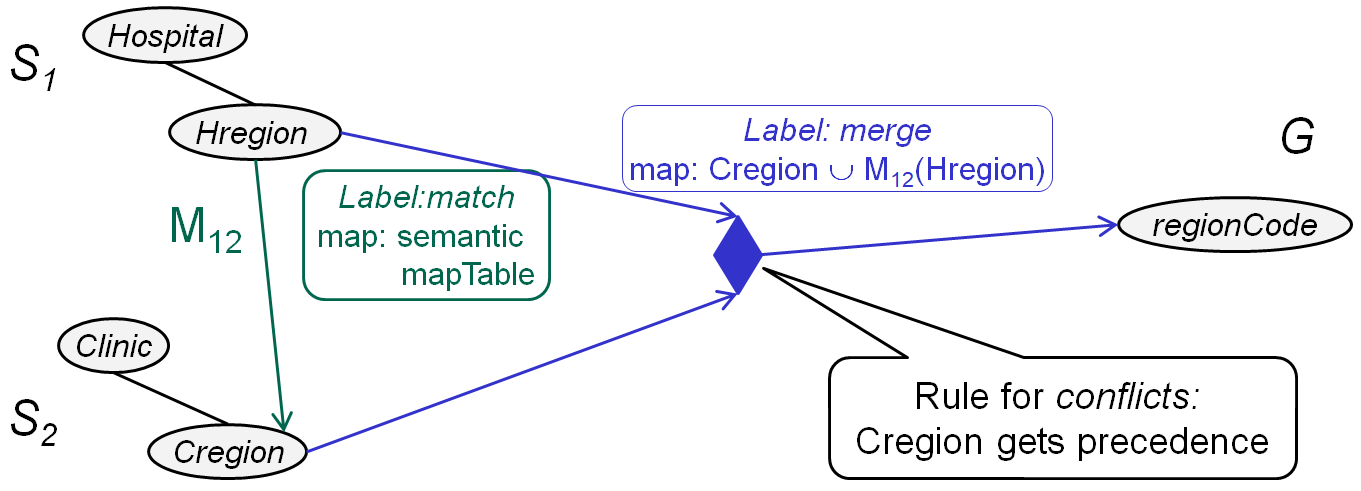}
\caption{Merge example of two Patient stats sources (Hospital and Clinic) with map table and conflict resolution}
\label{fig:Merge}
\end{figure}

Figure \ref{fig:Merge} shows a typical merge situation of two sources (Hospital and Clinic) with  different region coding. 
The mapping $M_{12}$ between Hregion and Cregion is required and allows to merge the overlapping data. 
In case of a value conflict the Cregion gets precedence. 

\subsubsection{Homomorphism Pattern}
A Homomorphism is a structure preserving mapping that helps to transform the source schemata into a target schema and thereby solves Problem 4) from the Introduction. 
It preserves edges but allows multiple nodes to map to the same target node. 
This can be used for data aggregation. 
If the mapping is injective (one-to-one), we have an Isomorphism. It transforms the source schemata into an equivalent target schema.

\begin{figure}[]
\centering
\includegraphics[width=0.48\textwidth]{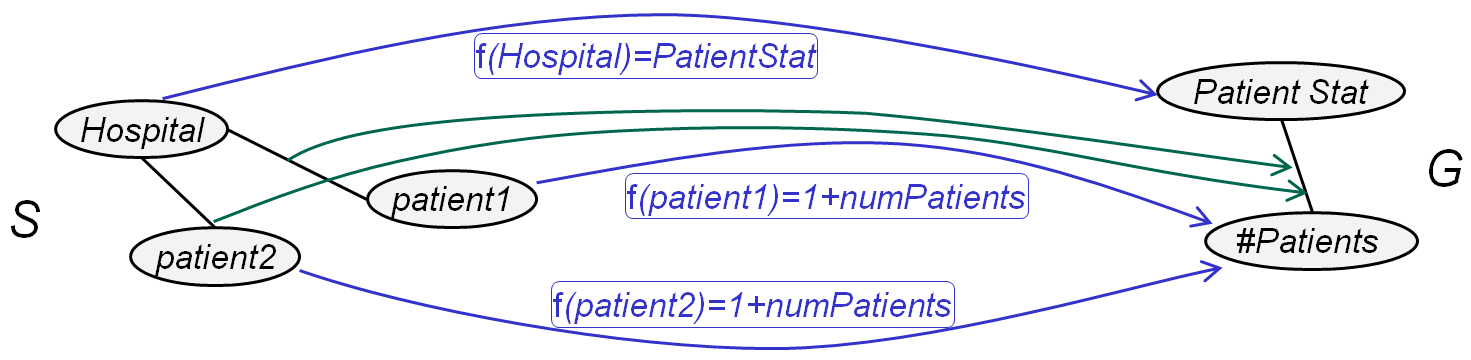}
\caption{Example instance graph Homomorphism that sums patients}
\label{fig:Homomorphism}
\end{figure}

In Figure \ref{fig:Homomorphism} the homomorphism f is a mapping that transforms all nodes and edges from schema $S$ onto nodes and edges of target $G$. In this example the patient nodes are mapped to the numPatients node of the target by incrementing the numPatients value. The edges between patient$x$ ($x = 1, 2$) and Hospital are mapped (green arrows) to the same edge (PatientStats--NumPatients).

\subsubsection{Commutative Mapping Pattern}
\begin{figure}[]
\centering
\includegraphics[width=0.48\textwidth]{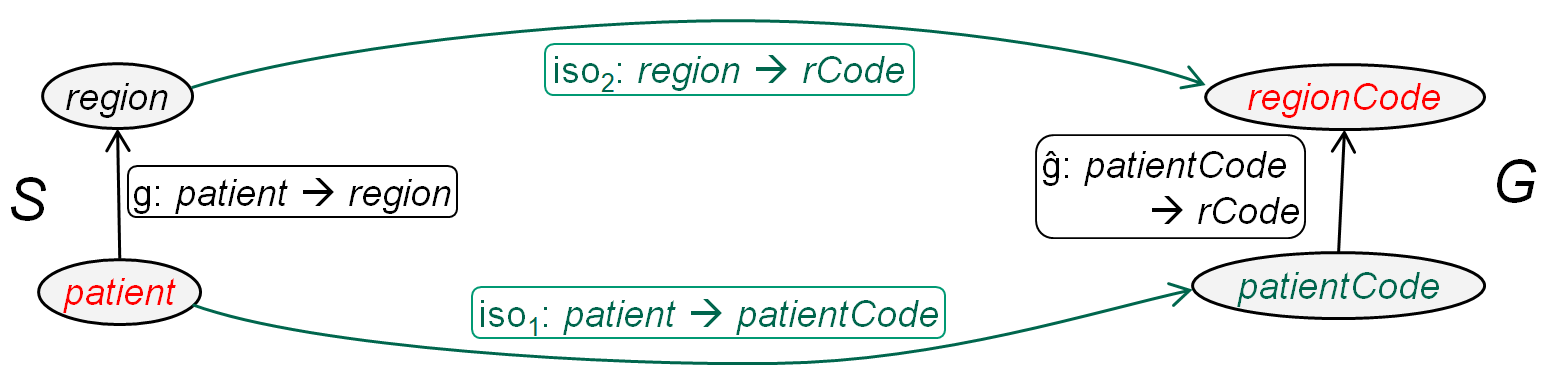}
\caption{Commutative mapping from patient to regionCode}
\label{fig:Commutative}
\end{figure}

When defining the mappings between two schemata special care has to be taken if a target node can be reached via different paths.
This happens for example when in the source schema two data items are related and both items are mapped to the same target node. 
In this case, we have to use the Merge pattern to resolve the conflict. 
But, if we preserve edges like in our example in Figure \ref{fig:Commutative}, we should have mappings that commute. 
A function chain is called commutative if and only if the order of the functions does not matter, i.e., $f_2 \circ f_1 = f_1 \circ f_2$. 
If special mappings are used like projection $g$ and isomorphism $iso$ then chances are good that the mapping chain is commutative. 
Communicative mappings are an essential criterion for a consistent mapping, and it helps to solve problem 5) from the Introduction.
Commutativity is important because it guarantees that we have the choice between alternative mapping paths and still end up with the same target data. In Figure \ref{fig:Commutative} it is irrelevant if the projection g to region is done first or the isomorphic mapping $iso_1$ to patientCode. We always end up with the same regionCode. 

\section{Quality Criteria (Task 4)}
\label{sec:Quality}

A bipartite graph is a graph where the nodes are separated in two disjoint subsets with no edges within the subsets.
This is the case with graph matching if we remove (in our mind) the connections inside the local schemata.
If we consider such a bipartite graph, we can define the following quality criteria:
A matching between the source and target graphs is called \emph{maximum matching} if the number of matched vertices is maximized.
If all nodes are matched, we call this a \emph{perfect matching}. 
This guarantees that all nodes (data elements) from the sources are matched with target nodes and this gives us a criterion of how well the matching covers the integration task.

The theorem of Hall \cite{Hall} states that there exists a perfect matching if all possible subsets of the source nodes have at least as many links to target nodes as the cardinality of this target subset.
More formally, let $V = (S \cup G, E)$ be a bipartite graph of two disjoint sets $S, G$, then there exists a perfect matching if $\forall R \subseteq S $ the inequality $d(R) \geq | R |$ holds where $d(R) := |\lbrace g \in G\mid r \in R \wedge (r,g) \in E\rbrace|$ is the number of nodes in $G$ linked to $R$.
The perfect matching is a general criterion for the coverage or completeness of a data integration. 
If no perfect match exists a merge conflict can arise and conflict resolution is necessary.

The theorem of Hall is only a formal quality criterion. In order to improve the semantic mapping quality, we may distinguish three cases on the instance level:
\begin{description}
\item[(1--1)] Data are mapped 1--1 to compatible data types or via an enumeration list. This may include disjoint merge operations.
\item[(n--1)] Data are mapped to an aggregated value or a merge with redundant data.
\item[(1--n)] Data are distributed to multiple data elements. This can occur for address data that is split into separate fields or split up values that include tax.
\end{description}

A semantic mapping quality measure can be established by assigning 3 points to every 1--1 mapping, 2 points to every n--1 mapping. 
The n--1 mapping loses information compared to the 1--1 mapping; therefore, it receives a lower score. 
The 1--n mappings receive 1 to 3 points depending on the reliability of the split operation.
Integration mappings with the highest sum represent the best match.
There exist many other schema quality measures for completeness, correctness, minimality, etc.
A nice overview on these measures is given in \cite{Ehrlinger}. 
These measures are of great value to quantify the quality of a schema but give no direct help how to decide between mapping options. 

Let us take Figure \ref{fig:Commutative} as example and calculate the quality measure for the pictured mappings. The 1--1 mappings $id_1$, $id_2$, $iso_1$, and $iso_2$ have 3 points each and the projections $g$ and $\hat g$ get 2 points each because of the n--1 mapping. 
The mapping chains $g \circ iso_2$ and $iso_1 \circ \hat g$ are commutative and yield the same score 5 (= 3 + 2), which confirms the equivalence of both mapping paths.

The matching and mapping in Figure \ref{fig:Mapping} also show a pair of commutative mappings from \textit{Hospital patient stats} to \textit{regionCode} with equal quality scores (see arrows with green glow, \emph{iso} = 3 points, projection $\pi$ = 2 points). If the scores differ, it is recommended to only use the mapping with the higher score because it represents a higher mapping quality.

\section{Conclusion and Future Work}
This paper presented an information integration process using the TGM with emphasis on semantically high quality as required for enterprise or government applications. 
Due to the TGS with predefined and user-defined data types, the TGM improves the formal data quality compared to other data integration approaches.
Integration patterns and quality criteria give guidelines for practical use of the matching and mapping task. 
The whole process was illustrated by a running example.
   
We conclude that high-quality integration with an engineered target schema is feasible. 
Some integration steps may be supported by automatic methods but still require human support with meta-data and context knowledge to achieve high-quality results.
The development of software to support schema generation as well as the matching and mapping remains as future work. 
In particular, the current expert-based mappings in Task 3 could be enhanced with semi-automated suggestions to the user with possible mapping options. 
One approach for such a program could use some elements from the academic prototypes G\textsc{RADOOP} \cite{Junghanns2015}, IncMap \cite{Pinkel} or Pregel \cite{Malewicz} to reduce the development effort.

\end{document}